\documentclass{aadebug}
\usepackage {ag-debug}
\corr{0309027}{283}
\input {epsf}
\usepackage{color}

\usepackage[dvips]{hyperref}
\makeatletter
\renewcommand{\section}{\@startsection{section}{1}%
{\z@}%
{-0.5ex \@plus -1ex \@minus -.1ex}%
{0.5ex \@plus.1ex}%
{\normalfont\Large\bfseries}}
\renewcommand{\subsection}{\@startsection{subsection}{1}%
{\z@}%
{-0.5ex \@plus -1ex \@minus -.1ex}%
{0.5ex \@plus.1ex}%
{\normalfont\large\bfseries}}
\makeatother
\begin {document} 
\runningheads{Sasaki et al.}{Generalized Systematic Debugging for Attribute Grammars}
\twocolumn[
\title{Generalized Systematic Debugging for Attribute Grammars}

\author{%
Akira Sasaki\addressnum{1}\comma\extranum{1},
Masataka Sassa\addressnum{2}\comma\extranum{2}
}
\address{1}{
The Advanced Clinical Research Center, The Institute of 
Medical Science, The University of Tokyo, Japan
}
\address{2}{
Department of Mathematical and Computing Sciences, Tokyo Institute of
Technology, Japan
}

\begin{abstract}
Attribute grammars (AGs) are known to be a useful formalism for semantic
analysis and translation. However, debugging AGs is complex owing to
inherent difficulties of AGs, such as recursive grammar structure
and attribute dependency. In this paper, a new systematic method of debugging
AGs is proposed. Our approach is, in principle, based on previously proposed
algorithmic debugging of AGs, but is more general.
This easily enables integration of various query-based systematic
debugging methods, including the slice-based method.  The proposed method 
has been implemented in Aki, a debugger for AG description. We evaluated
our new approach experimentally using Aki, which demonstrates the usability of
our debugging method.
\end{abstract}

]

\extra{1}{sasaki@ims.u-tokyo.ac.jp}
\extra{2}{sassa@is.titech.ac.jp}

\section{Introduction}
Debugging of attribute grammars (AGs) involves specific hurdles
because of the language features of AGs, such as recursion of syntax structure and
complex dependency between attributes. 
To attack these problems our group has developed two AG debugging
methods, one applying algorithmic debugging \cite{SassaOokubo97}
and the other with a slice-based debugging method for AGs
\cite{IkezoeSasaki2000}, which works in a complementary manner with the algorithmic
debugging-based method. 

Although the previous methods are effective for debugging AGs,
some limitations still exist. For example, the user may have to answer
questions about a huge tree, depending on the location of the bug.
Another limitation is that the user has no way to give
information directly to the debugger other than by answering a question from
the debugger. The most obvious problem of the previous methods is that
they work independently of each other---that is, the user cannot
switch to another method during debugging.

In this paper, we propose an AG debugging method that solves the abovementioned problems.
Our approach is a generalization of algorithmic
debugging of AGs \cite{SassaOokubo97}.  Whereas queries performed by the
previous methods have a single form, the new method allows several forms of
query. This enables integration of
various query based methods, including the previous two methods, in a
single framework.  We implemented the new debugging method in our
debugger. We showed the effectiveness of the proposed method 
experimentally.

\section{Algorithmic Debugging of\\ AGs}\label{sec: ad}
\emph{Attribute grammar} is a formalization that integrates both syntax and
semantics of languages. Fig. \ref{fig: grammar} is a simple example of
attribute grammar description. This description calculates the value of
a number in binary notation (including a bug).
\begin{figure}[h]
\begin{tabbing}
 F \= \ ::= \ \ \ . \ \ L \ \hspace{1.3cm}        \= L$_0$   \ ::=     \ \ B \ \ L$_1$ \\
\> \{ \attr L{}.{pos} = 1;                         \> \ \{ \attr L1.{pos} = \attr L0.{pos} + 1; \\
\> \ \ \attr F{}.{val} = \attr L{}.{val} \}        \> \ \ \ \attr B{}.{pos} = \attr L0.{pos} + 1; \ (bug) \\
 B  \>::= \ \ \  1                                 \> \ \ \ \attr L0.{val} = \attr B{}.{val} + \attr L1.{val} \} \\
\> \{ \attr B{}.{val} = $2^{-\attr B{}.{pos}}$ \}  \> \ \ $\mid$ \ \  B \\ 
\> \ \ $\mid$ \ \ 0                                \> \ \{ \attr B{}.{pos} = \attr L0.{pos}; \\
\> \{ \attr B{}.{val} = 0 \}                       \> \ \ \ \attr L0.{val} = \attr B{}.{val} \}
\end{tabbing}
  \caption{Attribute Grammar $G1$}
  \label{fig: grammar}
\end{figure}

\begin{figure}[h]
  \centerline{\includegraphics[width=.7\linewidth]{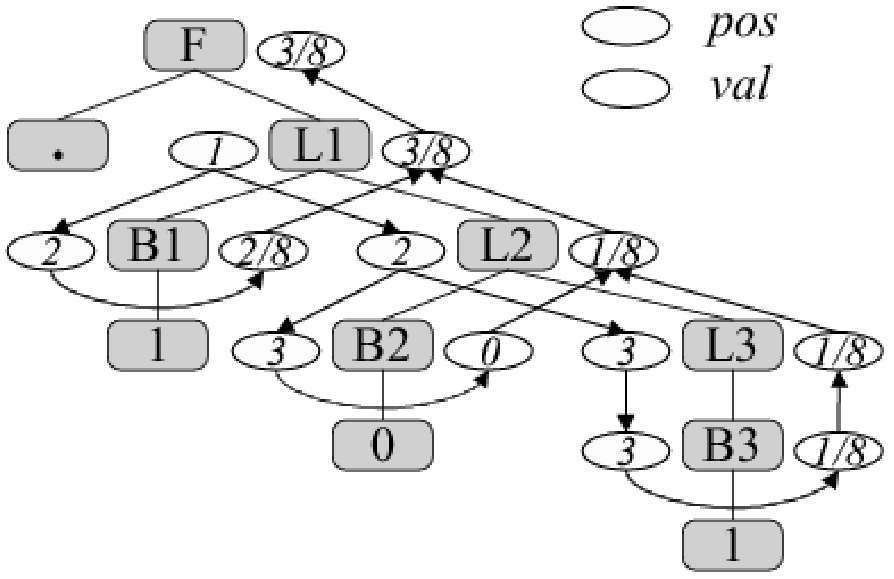}}
\caption{Attributed parse tree} 
\label{fig: apt}
\end{figure}

\emph{Attribute evaluation} is a process that calculates semantics
according to the attribute grammar.  
For $G1$, the attribute evaluation of ``.101'' is performed as follows:
(1) construction of the parse tree for input ``.101'', then
(2) computation of the value of each attribute according to the attribute dependency.
By this process, an attributed parse tree is constructed, as shown in Fig. \ref{fig: apt}. $F.val=3/8$ is the result of the evaluation, which is incorrect.

\begin{figure}[hb]
  \centerline{\includegraphics[width=.6\columnwidth]{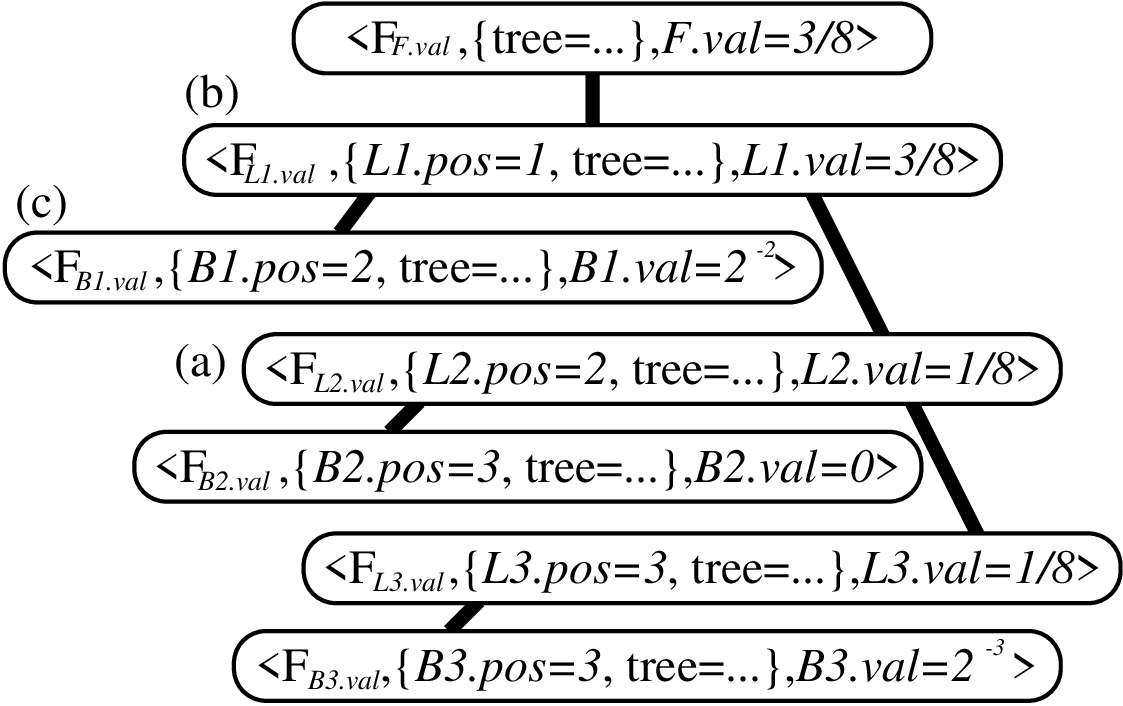}}
  \caption{Computation tree}
  \label{fig: comp-tree}
\end{figure}

\emph{Algorithmic debugging} proposed by Shapiro \cite{Shapiro82} is
formalized by \emph{computation trees}. A
computation tree represents the trace of program execution that
corresponds to a proof tree for logic languages. 
To apply this method to AGs, we need a structure equivalent
to the computation tree in AGs. Using fictitious functions called 
\emph{synth-functions} \cite{SassaOokubo97}, we can do this.  By
recursive application of synth-functions, we obtain a structure
equivalent to the computation tree that can model execution in
AGs, i.e. attribute evaluation.
Fig. \ref{fig: comp-tree} represents the computation
tree formed by the attribute evaluation of ``.101'' for $G1$.
Each node of the computation tree consists of a
triplet including a function name (of a synth-function), arguments and the result. 
For example, node (a) represents the computation 
that the value of $L2.val$ is 1/8 when $L2.pos=2$ and the
parse tree rooted at $L2$ (substring ``01'' of the input) are given. 
The user can confirm that this computation (relation) is correct and the
debugger can prune the tree rooted at (a) from the search space.

\section{Debugging Algorithm}\label{sec: gadimp}
\subsection{Problems with the previous approach}
\begin{figure}[htb]
  \centerline{\includegraphics[width=.45\columnwidth]{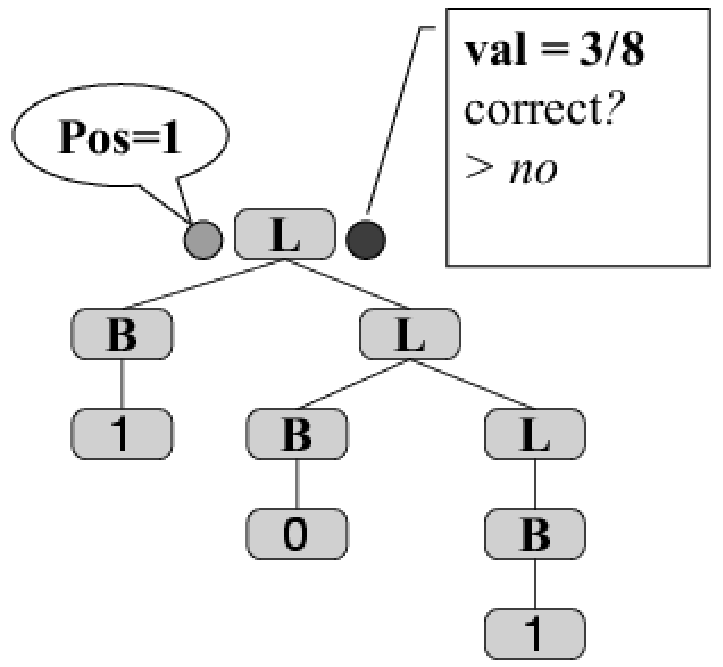}}
  \caption{Query with a large subtree} 
\label{fig: query1}
\end{figure}

There are difficulties with the previous algorithmic debugging of AGs
\cite{SassaOokubo97}.
One problem is that it is hard for the user to answer a question
near the root of the computation tree because the user requires 
information from the large subtrees (e.g. Fig. \ref{fig: query1}).

Another problem is the limitation on flexible debugging. We have developed
another systematic debugging for AGs, which is based on partition of
program slicing \cite{IkezoeSasaki2000}. However, after the user starts
debugging using program slicing, he or she generally must use that
method throughout the remainder of the debugging process, and
cannot switch to the algorithmic debugging (and vice versa).
Other problems exist: the previous algorithmic debugging is hard to
apply to a program that leads to run-time errors,
and algorithmic debugging generally may only indicate a plurality
of semantic rules as candidates for the bug.

\subsection{Debugging Algorithm}
As shown above, the sole use of the previous algorithmic debugging of
AGs is not effective. To solve these problems, we have developed a 
generalized version of algorithmic debugging for AGs, which enables
integration of various query based debugging methods, including slice-based
debugging. Compared with the method of Kamkar et al. \cite{Kamkar90},
which employs a combination of interprocedural slicing and algorithmic
debugging, we aim at a more generalized method for AGs that allows
users to switch between various debugging methods.

We have formalized several theorems for the generalization. Details of
these theorems are presented in \cite{SasakiSassa2003}. 

Fig. \ref{fig: gadalg} is a summary of the debugging method.
This algorithm localizes a bug using recursive functions:
$GAD_{init}$ is the function that gives
an initial condition, and $GAD$ is the actual function
that performs the recursive applications of the bug localization process.
The argument $\ACC$ of $GAD$ represents
an attribute computation composition the behavior of which is incorrect.
Here an ``attribute computation composition'' in a general sense means a
sub-computation in the attribute evaluation.
The second argument $\{\ACC_1,$ $\cdots,$ $\ACC_n$$\}$ represents a set of
attribute computation compositions the behaviors of which are correct.

\begin{figure}[htb]
\input{fig/gad}
\caption{Algorithm $GAD$}\label{fig: gadalg}
\end{figure}

Function $getNextACC$ selects an $\ACC'$ and $m$ for the next query.
Here we should select $\ACC'$ and $m$ that are subject to certain
properties: $\ACC'$ should include $\ACC_i$ $(1 \leq i \leq m)$, and $\ACC'$ should
not have an intersection with $\ACC_i$$(m < i)$. This function determines the
form of query to the user. That is, by changing the realization of
$getNextACC$, various debugging methods can be induced in this
algorithm, which gives the debugger flexibility. 

\begin{figure}[htb]
    \centerline{\includegraphics[width=.6\columnwidth]{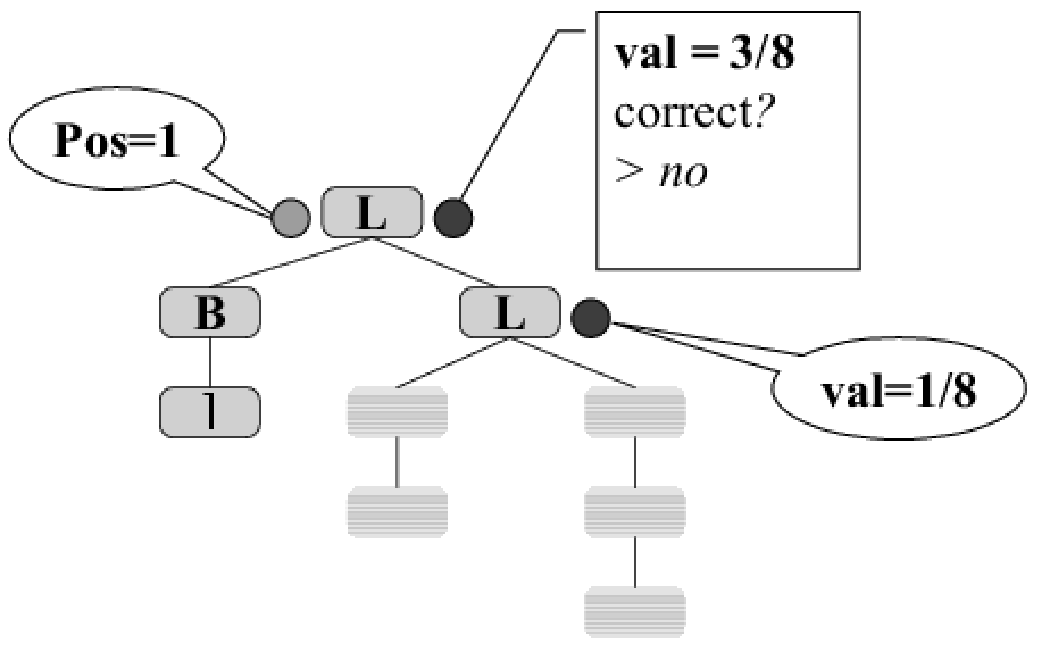}}
  \caption{Query with an incomplete subtree}
  \label{fig: query2}
\end{figure}

If \ACC' takes a form equivalent to a synth-function, 
the algorithmic debugging above can be realized. Slice-based debugging
can also be realized by selecting \ACC' that is equivalent to a
program slice---that is, sub-computations ranging from the start of the
attribute evaluation to some execution point.
To resolve the problem of huge trees (Fig.
\ref{fig: query1}), we can select \ACC' that generates a query such as
Fig. \ref{fig: query2} that is not a form of synth-functions.  
Therefore, we can easily realize and integrate
several debugging methods in a single framework.  

\section{Debugger Aki}\label{sec: aki}
We have implemented the debugger known as Aki \cite{IkezoeSasaki2000}, as a part of
our compiler development environment with AGs. In this environment, each
phase of the compiler is described in AG.  The debugger Aki assists
debugging of the descriptions of the compiler phases written in AGs.

Aki provided two debugging methods: the previous (i.e. naive)
algorithmic debugging and a method based on the partition of slices.
In addition, the generalized method in Section \ref{sec: gadimp} is
implemented for this research.

Aki performs systematic debugging using information
on attribute dependency that can be obtained from the
AG description, an input parse tree, and the trace of the evaluation of
attribute values.  To help the user understand its questions,
Aki implements several mechanisms. For example, for questions concerning
values with large data structures (such as symbol tables), Aki highlights the
difference between the two values, which helps the user to 
understand the value intuitively. 

\begin{figure}[tb]
  \centerline{\includegraphics[width=\columnwidth]{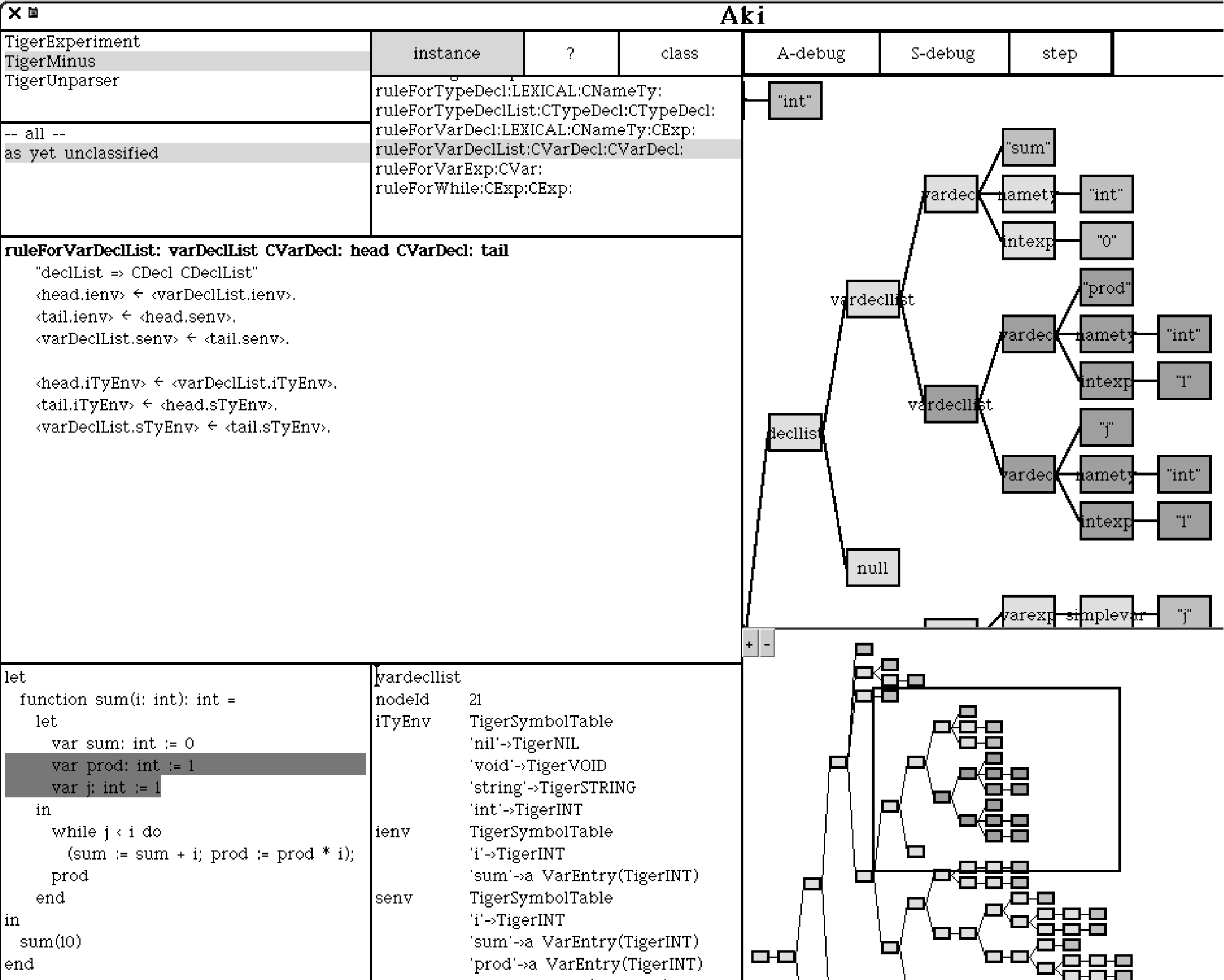}}
  \caption{Debugger Aki}
  \label{fig: aki}
\end{figure}

\begin{figure}[hbt]
  \centerline{\includegraphics[width=.6\linewidth]{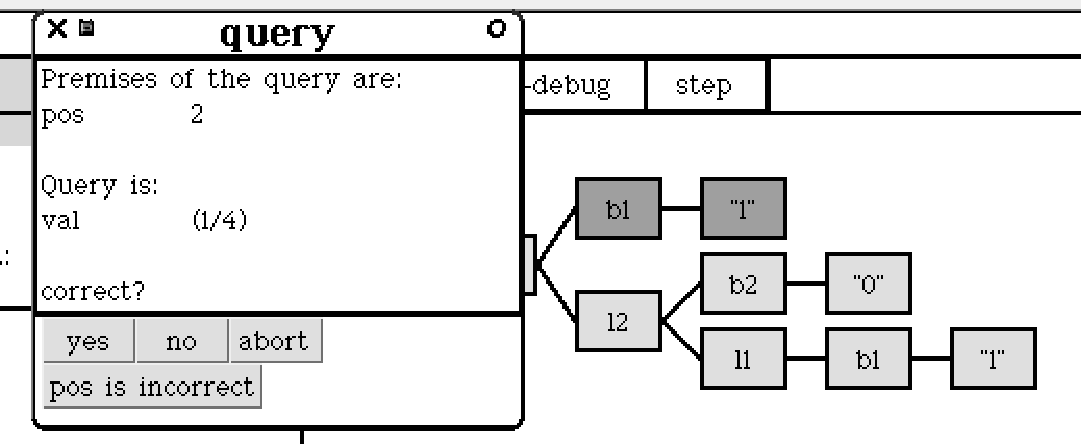}}
  \caption{Query}
  \label{fig: aki-query1}
\end{figure}

\begin{figure}[hbt]
  \centerline{\includegraphics[width=.7\linewidth]{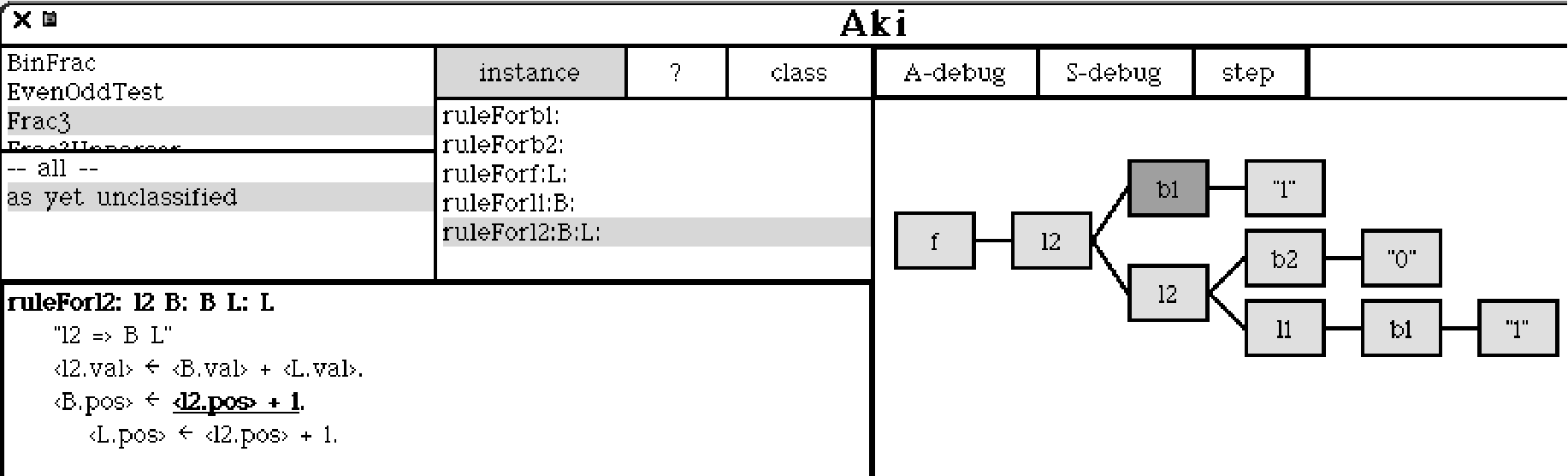}}
  \caption{An erroneous rule inferred by Aki}
  \label{fig: aki-query2}
\end{figure}

Fig. \ref{fig: aki} shows a screen shot of debugging
of an AG description using Aki. The panes display
the AG description, the source program, attribute values
and the input tree. Queries are presented in a dialog window,
as shown in Fig. \ref{fig: aki-query1}. When Aki detects
a semantic rule that includes a bug, Aki highlights the
rule as shown in Fig. \ref{fig: aki-query2}. 


\section{Experimental Results}
We performed user tests of the debugger Aki presented in the previous section.
Three test users tried to find erroneous positions in six programs using three
methods---i.e., slice-based, pure algorithmic and our
generalized method. 
The example attribute grammar is a description of static semantics
checking of the ``Tiger language'' in a compiler text \cite{Appel98}.
The description consists of 25 productions and 105 semantic rules.
The three users were all familiar with the language.
\begin{table}[h]
\caption{Comparison of the number of queries}\label{tab: result}
\begin{center}
\begin{tabular}{|c|r|r|r|r|l|}
\hline
& \# attrs  & \# nds & Slice & AD & GAD \\
\hline
$A$ & 78  &52	&7(1)	&8(3)	&4(1)[6(1)]\\ \hline 
$B$ & 146 &103	&6(1)	&9(4)	&5(1)\\ \hline 
$C$ & 60  &43	&4(1)	&8(3)	&4(1) \\ \hline 
$D$ & 56  &34     &9(1)	&6(4)	&7(1)\\ \hline \hline
$E$ & 45  &32	&5(1)	&5(2)	&7(2)\\ \hline 
$F$ &104  &69	&7(1)	&6(2)	&2(2)\\ \hline 
\end{tabular}
\end{center}
\end{table}

Table \ref{tab: result} shows the number of the queries
that the users required to identify
an erroneous portion in each description. In this table, ``\# attrs''
and ``\# nds'' denote the number of attribute instances and parse tree nodes,
respectively. ``Slice'',``AD'' and ``GAD'' represent the slice-based method,
pure algorithmic debugging and the method proposed in this paper,
respectively. Rows E and F are the cases for runtime error. The
numbers in parentheses are the numbers of candidates of semantic rules
identified as bugs. 

In this experiment, we could not conclude that a method with fewer queries is
more efficient, because some questions are difficult to answer and others
are easy. We discuss this in the next section. In the table,
the numbers in brackets in row A mean that one user had a different
result from other users. This is because users can freely indicate
erroneous values of inherited attributes in GAD.
\section{Discussion}
We discuss the advantages of the proposed debugging method. Some features
can be realized in the previous methods by ad-hoc extension. However, our
method is advantageous because these features can be realized in a
single framework.

\emph{Easy question to answer:}
In the previous two methods, to answer the question is the only way 
the user can give debuggers information for bug localization. On the
other hand, in the proposed method, the user can indicate some attribute
values that are found to be wrong. This enables the focus of the search
to be nearer the user's interest, which leads to efficient debugging.

\emph{Runtime error:}
When attribute evaluation produces a runtime error,
the previous algorithmic debugging of AGs forces the user to answer such difficult
question as, ``is it correct that an attribute value should be
undefined for this premise?''.  In the proposed method, the debugger
never asks a question including undefined attributes. This can be
realized by the integration of both pure algorithmic debugging
and slice-based debugging.

\emph{Reduce questions on big trees:}
In the previous algorithmic debugging of AGs, a subtree that is a premise
of a query should be a complete subtree with all descendant nodes.
On the other hand, in the proposed method, trees in a query may be
incomplete, in the sense that some part of the subtrees
may be pruned (e.g. in Fig. \ref{fig: query2}).

\emph{Identification of bug:}
When employing the previous algorithmic debugging of AGs, semantic rules
identified as erroneous are not, in general, single rules.  On the other
hand, the proposed method can identify just one semantic rule as
erroneous. 

\section{Concluding Remarks}
This paper presents a systematic debugging method for AGs.
We generalized the algorithmic debugging
of AGs, which allows various forms of questions and unifies the
previous two methods. We also developed a new debugging
method using the proposed framework. This method is
implemented in the debugger Aki, and experimental results are shown.

In the future we intend to develop a more effective method using the
generalized algorithmic debugging proposed in this paper.
We will also investigate the combination of the
proposed method and other debugging methods, such as assertion, as well as the
user interface of the debugger for questions that are easier to answer.

\end {document}